\documentstyle{article}

\title{Disentangling violations of CPT \\
from other new-physics effects}

\author{L.\ Lavoura$^a$ and Jo\~ao P.\ Silva$^{b,c}$\\
\\
\small $^a \!\!$ 
\small Universidade T\'ecnica de Lisboa \\
\small Centro de F\'\i sica das Interac\c c\~oes Fundamentais \\
\small Instituto Superior T\'ecnico, 1049-001 Lisboa, Portugal \\
\small Email i009@beta.ist.utl.pt.
Telephone \raise.25ex\hbox{\footnotesize ++}351 1 8419093.
Fax \raise.25ex\hbox{\footnotesize ++}351 1 8419143. \\
\\
\small $^b \!\!$ Centro de F\'\i sica Nuclear da Universidade de Lisboa \\
\small Avenida Professor Gama Pinto, 2 \\
\small 1699 Lisboa Codex, Portugal \\
\\
\small $^c \!\!$ Centro de F\'\i sica \\
\small Instituto Superior de Engenharia de Lisboa \\
\small 1900 Lisboa, Portugal}

\date{15 February 1999}

\begin{document}
\maketitle

\begin{abstract}
We analyze the prospects for observing CPT violation
using neutral-meson $P^0$-$\overline{P^0}$ systems.
Before one can claim a measurement of CPT violation,
one must be able to rule out the possibility
that its result is due to simpler new-physics effects.
In particular,
one must be able to separate the CPT-violating quantities
from parameters violating the $\Delta P = \Delta Q$ rule
of semileptonic decays,
and from new-physics contributions to the production mechanism
of the neutral mesons.
One can isolate CPT violation using the semileptonic decays of single,
tagged neutral mesons;
unfortunately,
this situation cannot be implemented at the $\Upsilon(4S)$.
For $P^0 \overline{P^0}$ pairs produced
in a correlated parity-odd state
we show that,
by combining the di-lepton with the single-lepton decays,
it is in principle possible to extract unambiguously
one CPT-violating parameter.
Finally,
we develop the formalism necessary for describing new-physics effects
in the production mechanism;
this includes both cascade decays and
violations of the rule of associated production.
\end{abstract}

\section{Introduction}

The ``CPT theorem'' states that any local field theory with hermitian,
Lorentz-invariant interactions
obeying the spin-statistics connection
is necessarily CPT-invariant \cite{proofs}.
Although the assumptions of this theorem---and thus
the validity of its conclusion---are generally taken for granted,
the question of whether CPT is violated or not
should ultimately be settled through accurate,
high-precision experiments.

The most elementary consequences of CPT invariance
are the equal values of the masses and lifetimes,
and the symmetrical values of the magnetic moments,
of a particle and its antiparticle.
Unfortunately,
in these cases the prospective CPT violation is expected
to be a small perturbation in quantities which are dominated by 
much stronger interactions.
For instance,
the difference between the masses of $K^+$ and $K^-$
has an experimental bound \cite{PDG}
$\left| m_{K^+} - m_{K^-} \right| / \left( m_{K^+} + m_{K^-} \right)$
\raise.3ex\hbox{$<$\kern-.75em\lower1ex\hbox{$\sim$}} $10^{-4}$;
however,
this bound is not very meaningful,
since $m_{K^+}$ and $m_{K^-}$
are dominated by the strong interaction,
while one would expect CPT violation to be at best
of miliweak or even superweak strength.

In contrast,
the mixing between a neutral meson $P^0$ and its antiparticle $\overline{P^0}$
(here $P^0$ may be either $K^0$,
$D^0$,
$B^0_d$,
or $B^0_s$)
is a second-order electroweak effect.
The smallness of the mixing makes it an ideal setting
to look for small violations of the symmetries CP,
T,
and CPT.
In fact,
CP violation was first established \cite{Chr64}
in the $K^0$-$\overline{K^0}$ system,
and has thus far eluded experimental detection in any other system.
Recently,
the CPLEAR Collaboration has presented the results
of its search for T violation \cite{CPLEAR-T}
and CPT violation \cite{CPLEAR-CPT}
in $K^0$-$\overline{K^0}$ mixing,
and the OPAL Collaboration has looked for CPT-violating effects
in the $B^0_d$-$\overline{B^0_d}$ system \cite{OPAL}.
Detailed experiments on the $K^0$-$\overline{K^0}$
and $B^0_d$-$\overline{B^0_d}$ systems
are planned at the $\phi$- and $\Upsilon(4S)$-factories,
respectively.

Most of these experiments involve one or both of the following crucial steps:
1) determining the flavor of the initial meson
(this procedure is called ``tagging'');
2) determining the flavor of the meson at decay time,
which is usually done by looking for semileptonic final states.
When one is searching for CPT violation one must face the possibility that
both steps are affected by new physics;
one must make sure that what is assumed to be a measurement of CPT violation
is not,
in reality,
a measurement of a much less revolutionary new-physics effect.

The semileptonic decays of the mesons obey,
in the standard model (SM) and to first order in the electroweak interaction,
the $\Delta P = \Delta Q$ rule
($Q$ is the hadrons' charge,
and $P$ the flavor quantum number of the heaviest quark in the decaying meson).
That is,
in the SM,
the decays $P^0 \rightarrow X^- l^+ \nu_l$ and
$\overline{P^0} \rightarrow X^+ l^- \bar \nu_l$ are allowed,
while the decays $P^0 \rightarrow X^+ l^- \bar \nu_l$ and 
$\overline{P^0} \rightarrow X^- l^+ \nu_l$
(which have $\Delta P = - \Delta Q$)
are only possible at second order in the electroweak interaction.
(Here,
$X^\pm$ denotes a pair of arbitrary CP-conjugate hadronic states.)
It is usual to define the parameters
\begin{eqnarray}
x_l &=& \frac{\langle X^- l^+ \nu_l | T | \overline{P^0} \rangle}
{\langle X^- l^+ \nu_l | T | P^0 \rangle},
\nonumber\\
\bar x_l &=& \frac{\langle X^+ l^- \bar \nu_l | T | P^0 \rangle^\ast}
{\langle X^+ l^- \bar \nu_l | T | \overline{P^0} \rangle^\ast}.
\label{xl-definition}
\end{eqnarray}
Their phases are not rephasing-invariant and,
therefore,
they are physically meaningless.
On the other hand,
the magnitudes of $x_l$ and of $\bar x_l$ are physically meaningful,
and they are generally assumed to be small.
If $x_l$ and $\bar x_l$ do not vanish,
then they will obscure the identification of CPT violation.

The tagging strategies used in most experiments assume that,
for any given event,
there is a clear signal of the initial flavor of the neutral meson.
The basic idea is rooted in the fact that the interactions of the gluon,
photon,
and $Z^0$ are flavor-conserving.
For example,
in the CPLEAR experiment \cite{CPLEAR-T,CPLEAR-CPT}
the neutral kaons are produced by the strong interaction
through the reactions $p \bar p \rightarrow K^- \pi^+ K^0$
and  $p \bar p \rightarrow K^+ \pi^- \overline{K^0}$;
the sign of the charge of the charged kaon identifies
the initial flavor of the associated neutral meson.
This is known as the rule of associated production.
Another example occurs at the $Z^0$ pole,
when $B^0$ is produced in association with $B^-$
and a set of particles with total charge $+1$.
Detecting the $B^-$,
either by reconstruction
or through its subsequent semileptonic decay---assuming
the $\Delta B = \Delta Q$ rule---one tags
the initial flavor of the neutral meson.
This was the strategy followed by the OPAL Collaboration \cite{OPAL}.
In these cases,
the production of the ``wrong'' neutral meson
would mean the existence of a $\left| \Delta P \right| = 2$ interaction.
However,
given that such an interaction would also contribute to 
$P^0$-$\overline{P^0}$ mixing,
its contribution to the production process
is usually assumed to be negligibly small.

This is no longer the case when the production process
is due to the  $\left| \Delta P \right| = 1$ interaction of the $W$ boson,
such as in cascade decays.
For example,
one may wish to study the $K^0$-$\overline{K^0}$ system
in the decay chain $B^0_d \rightarrow J/\psi K^0 \rightarrow J/\psi f$.
In the SM,
the analysis of this decay \cite{cascade} is based on the fact that
the decay $B^0_d \rightarrow J/\psi \overline{K^0}$
does not exist at tree level:
there are $\Delta B = - \Delta S$ transitions,
but no $\Delta B = \Delta S$ transitions. 
However,
new-physics effects might alter this situation.
One must be able to rule out such effects before
these processes can be used to look for violations of CPT.
Similarly \cite{Meca98},
one has access to the $D^0$-$\overline{D^0}$ system
through the decay chain 
$B^- \rightarrow K^- \{ D^0,\overline{D^0} \} \rightarrow K^- f$.
Here the situation is more complicated because both amplitudes
$B^- \rightarrow K^- D^0$ and $B^- \rightarrow K^- \overline{D^0}$ exist,
even within the SM \cite{Meca98,Amorim99}.

Measurements of CPT violation in tagged decays of neutral kaons
have been discussed in the literature,
sometimes including the possibility that the $\Delta S = \Delta Q$
rule is violated \cite{Sachs63,Dass72,Barmin84,Tanner86,Lavoura91}.
The subject has resurfaced in recent analyses \cite{Gaume98,Kabir99}
of the claim of an observation of T violation
by the CPLEAR Collaboration \cite{CPLEAR-T}.
The possibility of a measurement of CPT-violating effects
in the regeneration of neutral kaons has also been discussed
\cite{Lavoura91,Briere89}.
Measurements of CPT violation at the $\Upsilon(4S)$-factories
have been considered either assuming the $\Delta B = \Delta Q$ rule
\cite{Kobayashi92,Xing94,Mohapatra98},
or making some simplifying assumptions
about the nature of the $\Delta B = - \Delta Q$ amplitudes \cite{kostelecky}.
Conversely,
the experimental search for $\Delta S = - \Delta Q$ amplitudes
in neutral-kaon decays \cite{niebergall},
and the theoretical discussion
of a search for $\Delta B = - \Delta Q$ amplitudes
at the $\Upsilon(4S)$-factories \cite{Dass94},
have been made assuming CPT invariance.
Xing \cite{Xing98} has recently shown that it is impossible to disentangle
the violation of CPT from $\Delta B = - \Delta Q$ transitions
by using exclusively di-lepton decays of the $\Upsilon(4S)$;
we confirm his result here.

In this article we present a study
of the measurements of CPT violation enabled by semileptonic decays,
and of their impediment due to new-physics effects.
We consider for the first time
the impact of mis-taggings in the production process,
and we relate our results to Xing's conclusion
on the impossibility to disentangle CPT-violating amplitudes
from $\Delta B = - \Delta Q$ amplitudes in di-lepton events
at the $\Upsilon(4S)$.
However,
we also consider single-lepton events,
and we show that that separation is in principle possible
through a combination of single- and di-lepton events.
We stress that the situation described by tagged decays
cannot be implemented at the $\Upsilon(4S)$,
if we allow for violations of the $\Delta P = \Delta Q$ rule.

We define our notation in section 2.
In section 3 we discuss decays in which one has
tagged the initial flavor of the neutral meson
through the rule of associated production.
In section 4 we focus on neutral-meson pairs produced in a parity-odd state,
such as the $K^0 \overline{K^0}$ pairs from the decay of the $\phi$ resonance,
and the $B^0_d \overline{B^0_d}$ pairs produced at the $\Upsilon(4S)$.
In section 5 we show that the presence of new physics
in the production process will impede the extraction
of the CPT-violating parameters.
We draw our conclusions in section 6.

\section{Mixing formalism}

We start by discussing the mixing of $P^0$ and $\overline{P^0}$
in the Wigner--Weisskopf formalism,
when CPT violation is allowed for.
We do this in order to introduce a convenient parametrization
of the violation of T and CPT in the mixing,
and to establish our notation.
{\em All equations in this section are exact},
and we are careful to identify the reparametrization-invariant quantities;
only those quantities are physically meaningful.

The time evolution of
\begin{equation}
| \psi (t) \rangle = \psi_1 (t) | P^0 \rangle
+ \psi_2 (t) | \overline{P^0} \rangle
\label{the main state}
\end{equation}
is given by
\begin{equation}
i \frac{d}{dt}
\left( \begin{array}{c} \psi_1 (t) \\ \psi_2 (t) \end{array} \right) =
\left( \begin{array}{cc} R_{11} & R_{12} \\ R_{21} & R_{22} \end{array} \right)
\left( \begin{array}{c} \psi_1 (t) \\ \psi_2 (t) \end{array} \right).
\label{the main evolution}
\end{equation}
The eigenvalues of $R$ are denoted $\mu_a$ and $\mu_b$.
Their sum is given by the trace of $R$:
\begin{equation}
\mu_a + \mu_b = R_{11} + R_{22}.
\label{trace}
\end{equation}
The right-eigenvectors of $R$
corresponding to the eigenvalues $\mu_a$ and $\mu_b$
are $(p_a, q_a)^T$ and $(p_b, -q_b)^T$,
respectively:
\begin{eqnarray}
\left( \begin{array}{cc} R_{11} & R_{12} \\ R_{21} & R_{22} \end{array} \right)
\left( \begin{array}{c} p_a \\ q_a \end{array} \right)
&=& \mu_a
\left( \begin{array}{c} p_a \\ q_a \end{array} \right),
\nonumber\\
\left( \begin{array}{cc} R_{11} & R_{12} \\ R_{21} & R_{22} \end{array} \right)
\left( \begin{array}{c} p_b \\ - q_b \end{array} \right)
&=& \mu_b
\left( \begin{array}{c} p_b \\ - q_b \end{array} \right).
\label{eigenvectors}
\end{eqnarray}
Therefore,
\begin{eqnarray}
\frac{q_a}{p_a}
&=& \frac{\mu_a - R_{11}}{R_{12}}
= \frac{R_{21}}{\mu_a - R_{22}},
\nonumber\\
\frac{q_b}{p_b}
&=& \frac{R_{11} - \mu_b}{R_{12}}
= \frac{R_{21}}{R_{22} - \mu_b}.
\label{eigenvectors, explicit}
\end{eqnarray}
Equations~(\ref{trace}) and (\ref{eigenvectors, explicit}) imply
\begin{equation}
\theta =
\left( \frac{q_a}{p_a} - \frac{q_b}{p_b} \right)
/ \left( \frac{q_a}{p_a} + \frac{q_b}{p_b} \right)
= \frac{R_{22} - R_{11}}{\mu_a - \mu_b}.
\label{theta}
\end{equation}
In order to avoid using the three non-independent quantities $q_a / p_a$,
$q_b / p_b$,
and $\theta$,
it is convenient to introduce
\begin{equation}
\frac{q}{p} = \sqrt{\frac{q_a q_b}{p_a p_b}} = \sqrt{\frac{R_{21}}{R_{12}}}.
\label{ratio}
\end{equation}
Notice that we do not define the quantities $q$ and $p$ separately;
we only define the ratio $q/p$.
From Eqs.~(\ref{theta}) and (\ref{ratio}) it follows that
\begin{equation}
\sqrt{1 - \theta^2} = 2\, \frac{q}{p}\, /
\left( \frac{q_a}{p_a} + \frac{q_b}{p_b} \right).
\label{square root}
\end{equation}
The CPT-violating parameter $\theta$ will later be assumed to be small.
We shall then make
the approximation $\sqrt{1 - \theta^2} \approx 1$.

It follows from Eqs.~(\ref{the main state}),
(\ref{the main evolution}),
and (\ref{eigenvectors}) that the states
\begin{eqnarray}
| P_a \rangle &=& p_a | P^0 \rangle + q_a | \overline{P^0} \rangle,
\nonumber\\
| P_b \rangle &=& p_b | P^0 \rangle - q_b | \overline{P^0} \rangle
\label{ph and pl}
\end{eqnarray}
evolve in time as
\begin{eqnarray}
| P_a (t) \rangle &=& e^{- i \mu_a t} | P_a \rangle,
\nonumber\\
| P_b (t) \rangle &=& e^{- i \mu_b t} | P_b \rangle.
\label{primary evolution}
\end{eqnarray}
We do not have to make any assumption about the normalization
of $| P_a \rangle$ and of $| P_b \rangle$.
We also do not have to make any assumption either about the relative phase
of $| P_a \rangle$ and $| P_b \rangle$,
or about the relative phase
of $| P^0 \rangle$ and $| \overline{P^0} \rangle$.
Indeed,
one is free to change the phase
of the kets $| P^0 \rangle$ and $| \overline{P^0} \rangle$:
\begin{eqnarray}
| P^0 \rangle & \rightarrow & e^{i \gamma} | P^0 \rangle,
\nonumber\\
| \overline{P^0} \rangle & \rightarrow & e^{i \overline{\gamma}}
| \overline{P^0} \rangle.
\label{ket rephasing}
\end{eqnarray}
The invariance of the state vector $| \psi (t) \rangle$
under this rephasing implies that
\begin{eqnarray}
\psi_1 (t) & \rightarrow & e^{- i \gamma} \psi_1 (t),
\nonumber\\
\psi_2 (t) & \rightarrow & e^{- i \overline{\gamma}} \psi_2 (t).
\end{eqnarray}
Therefore,
from Eq.~(\ref{the main evolution}),
\begin{eqnarray}
R_{12} & \rightarrow & e^{i \left( \overline{\gamma} - \gamma \right)} R_{12},
\nonumber\\
R_{21} & \rightarrow & e^{i \left( \gamma - \overline{\gamma} \right)} R_{21},
\label{rephasing}
\end{eqnarray}
while $R_{11}$ and $R_{22}$ do not change.
The trace and the determinant of $R$
are invariant under the transformation in Eqs.~(\ref{rephasing}).
Therefore,
$\mu_a$ and $\mu_b$ are invariant too.
Thus,
$\theta$ is invariant under a rephasing of $| P^0 \rangle$
and $| \overline{P^0} \rangle$.
Both the real and the imaginary parts of $\theta$ are physically meaningful.
They violate CP and CPT.
On the contrary,
the phase of the parameter $q/p$ in Eq.~(\ref{ratio})
is not invariant under the rephasing in Eqs.~(\ref{rephasing});
as a result,
it is physically meaningless.
However,
the modulus of $q/p$ is physically meaningful;
the real parameter
\begin{equation}
\delta
=
%\frac{1 - \left| \frac{q}{p} \right|^2}{1 + \left| \frac{q}{p} \right|^2}
\left( 1 - \left| \frac{q}{p} \right|^2 \right)
/ \left( 1 + \left| \frac{q}{p} \right|^2 \right)
=
\frac{\left| R_{12} \right| - \left| R_{21} \right|}
{\left| R_{12} \right| + \left| R_{21} \right|}
\end{equation}
violates CP and T.

In summary,
the $P^0$-$\overline{P^0}$ mass matrix has two CP- and CPT-violating
parameters ($\mbox{Re}\, \theta$ and $\mbox{Im}\, \theta$),
and one CP- and T-violating parameter ($\delta$).
In addition,
it has four C-, P- and T-invariant quantities:
\begin{eqnarray}
m_a = \mbox{Re}\, \mu_a,
& \hspace{5mm} &
\Gamma_a = - 2\, \mbox{Im}\, \mu_a,
\nonumber\\
m_b = \mbox{Re}\, \mu_b,
& \hspace{5mm} &
\Gamma_b = - 2\, \mbox{Im}\, \mu_b.
\end{eqnarray}
These are sometimes traded for
\begin{equation}
m = \frac{m_a + m_b}{2},
\hspace{8mm}
\Gamma = \frac{\Gamma_a + \Gamma_b}{2},
\end{equation}
and
\begin{equation}
x = \frac{m_a - m_b}{\Gamma},
\hspace{8mm}
y = \frac{\Gamma_a - \Gamma_b}{2 \Gamma}.
\end{equation}

\section{Tagged decays}

Let us consider a neutral meson which is identified as a
$P^0$ ($\overline{P^0}$) at time $t=0$.
Using Eqs.~(\ref{theta})--(\ref{primary evolution}),
one finds that this state is given at time $t$ by
\begin{eqnarray}
| P^0 (t) \rangle &=& g_+ (t) | P^0 \rangle
+ g_- (t) \left( \frac{q}{p} \sqrt{1 - \theta^2} | \overline{P^0} \rangle
- \theta | P^0 \rangle \right),
\nonumber\\
| \overline{P^0} (t) \rangle &=& g_+ (t) | \overline{P^0} \rangle
+ g_- (t) \left( \frac{p}{q} \sqrt{1 - \theta^2} | P^0 \rangle
+ \theta | \overline{P^0} \rangle \right),
\label{evolution}
\end{eqnarray}
respectively.
Here,
\begin{equation}
g_\pm (t) = {\textstyle \frac{1}{2}}
\left( e^{- i \mu_a t} \pm e^{- i \mu_b t} \right).
\end{equation}

We now seek the quantities which can be measured when $| P^0 (t) \rangle$
and $| \overline{P^0} (t) \rangle$ decay into a final state $f$.
We define
\begin{equation}
A_f = \langle f | T | P^0 \rangle,
\hspace{5mm}
\bar A_f = \langle f | T | \overline{P^0} \rangle.
\label{the amplitudes}
\end{equation}
Equations~(\ref{evolution}) depend on two independent
functions of the decay time,
$g_+ (t)$ and $g_- (t)$.
Therefore,
one will in principle be able to measure the ratio of the coefficients
of the two functions,
\begin{eqnarray}
E &=& \frac{q}{p}\, \frac{\bar A_f}{A_f} \sqrt{1 - \theta^2}  - \theta,
\nonumber\\
\bar E &=& \frac{p}{q}\, \frac{A_f}{\bar A_f} \sqrt{1 - \theta^2} + \theta.
\label{tagged measurables}
\end{eqnarray}
We cannot compare the normalization of the decay rates corresponding
to different final states unless simplifying assumptions are made.
For example,
some authors assume that there are no electromagnetic final-state interactions,
or that CPT violation is absent in the decay process,
or that there is no T violation in the mixing of the neutral mesons. 
We would argue that all these effects must be considered
when looking for CPT violation,
which is in itself dramatically non-standard.

We stress that the observables in Eqs.~(\ref{tagged measurables})
contain {\em the maximal information} that may be extracted
from the time dependence of the decay rate.
It is possible that particular phenomenological or experimental conditions
only allow the extraction of part of this information
from the actual decay curves.
In order to see this,
consider the explicit decay rates:
\begin{eqnarray}
\Gamma [P^0 (t) \rightarrow f] &=& |A_f|^2 \left\{
\left| g_+(t) \right|^2 + \left| E \right|^2 \left| g_-(t) \right|^2 +
2\, \mbox{Re} \left[ E g_+^\ast(t) g_-(t) \right] \right\},
\nonumber\\
\Gamma[\overline{P^0} (t) \rightarrow f] &=& |\bar A_f|^2 \left\{
\left| g_+(t) \right|^2 + \left| \bar E \right|^2 \left| g_-(t) \right|^2 +
2\, \mbox{Re} \left[ \bar E g_+^\ast(t) g_-(t) \right] \right\}.
\hspace*{5mm}
\label{tagged decay rates}
\end{eqnarray}
Since the functions $\left| g_+(t) \right|^2$,
$\left| g_-(t) \right|^2$,
$\mbox{Re} \left[ g_+^\ast(t) g_-(t) \right]$,
and $\mbox{Im} \left[ g_+^\ast(t) g_-(t) \right]$
are linearly independent,
one can measure the quantities in Eqs.~(\ref{tagged measurables})
by tracing the time dependence of the decays of single,
tagged $P^0$ or $\overline{P^0}$.
However,
if for instance the two eigenstates have equal decay widths,
{\it i.e.},
if $\Gamma_a = \Gamma_b$,
then the function $\mbox{Re} \left[ g_+^\ast(t) g_-(t) \right]$
will vanish and less information will be available.
Thus,
in the ensuing discussions we address the best possible scenario.
Actual experiments may be considerably more problematic.

\subsection{Decays into semileptonic final states}

Consider the particular case of the semileptonic decays.
The parameters $x_l$ and $\bar x_l$ were defined in Eqs.~(\ref{xl-definition}).
We shall not need to assume any relationship
between $x_l$ and $\bar x_l$;
in particular,
we shall not assume CPT invariance of the decay amplitudes.
Now,
$x_l$ and $\bar x_l$ are not invariant under the rephasing of $| P^0 \rangle$
and $| \overline{P^0} \rangle$ in Eqs.~(\ref{ket rephasing}).
The rephasing-invariant,
physically meaningful quantities are
\begin{equation}
\lambda_l  =  \frac{q}{p}\, x_l,
\hspace{5mm}
\bar \lambda_l  =  \frac{p}{q}\, \bar x_l^\ast.
\end{equation}
They will be assumed to be small.
If one observes the tagged decays to the semileptonic state $X^- l^+ \nu_l$,
one can in principle measure the corresponding parameters $E$ and $\bar E$,
namely,
\begin{eqnarray}
\sqrt{1 - \theta^2} \lambda_l - \theta &\approx& \lambda_l - \theta,
\nonumber\\
\sqrt{1 - \theta^2} \left( \lambda_l \right)^{-1} + \theta
&\approx& \left( \lambda_l \right)^{-1}.
\label{tagged-l+}
\end{eqnarray}
If one observes the tagged decays to $X^+ l^- \bar \nu_l$,
one measures
\begin{eqnarray}
\sqrt{1 - \theta^2} \left( \bar \lambda_l \right)^{-1} - \theta
&\approx& \left( \bar \lambda_l \right)^{-1},
\nonumber\\
\sqrt{1 - \theta^2} \bar \lambda_l + \theta &\approx& \bar \lambda_l + \theta.
\label{tagged-l-}
\end{eqnarray}
In both Eqs.~(\ref{tagged-l+}) and (\ref{tagged-l-})
we have made the approximation of neglecting the products
of any two small parameters like $\theta$,
$\lambda_l$,
or $\bar \lambda_l$.
One sees that,
by using the decays of single tagged mesons,
one can in principle separate the CPT-violating parameter $\theta$
from the $\Delta P = - \Delta Q$ parameters $\lambda_l$ and $\bar \lambda_l$.
Thus,
{\em one can measure CPT violation with tagged decays}.

Unfortunately,
as will be shown in section 5.1,
this strategy cannot be implemented at the $\Upsilon(4S)$.

\subsection{About the searches for $\Delta P = - \Delta Q$ amplitudes}

The primary aim of this article is to stress the impact
that simple new-physics effects may have
on experiments seeking to measure violations of CPT.
We now want to point out that the converse is also true.
In particular,
the experiments performed in the 70's
in order to measure violations of the $\Delta S = \Delta Q$ rule
in the semileptonic decays of the neutral kaons
have disregarded the possibility of CPT violation.
Let us take $P^0$ to be $K^0$ and $\overline{P^0}$ to be $\overline{K^0}$.
For the subscripts which label the eigenstates of propagation
we use $a \rightarrow L$ and $b \rightarrow S$,
referring to the long-lived and to the short-lived neutral kaon,
respectively.
Denoting
\begin{eqnarray}
\lambda_e & = & \frac{q}{p}\, 
\frac{\langle \pi^- e^+ \nu | T | \overline{K^0} \rangle}
{\langle \pi^- e^+ \nu | T | K^0 \rangle},
\nonumber\\
\bar \lambda_e & = & \frac{p}{q}\, 
\frac{\langle \pi^+ e^- \nu | T | K^0 \rangle}
{\langle \pi^+ e^- \nu | T | \overline{K^0} \rangle},
\end{eqnarray}
one easily finds
\begin{eqnarray}
\Gamma [ K^0 (t) \rightarrow \pi^- e^+ \nu ] &=&
\frac{| \langle \pi^- e^+ \nu | T | K^0 \rangle |^2}{4}
\left|
\left( 1 - \sqrt{1 - \theta^2} \lambda_e + \theta \right) e^{- i \mu_S t}
\right.
\nonumber\\
& &
\left.
+ \left( 1 + \sqrt{1 - \theta^2} \lambda_e - \theta \right) e^{- i \mu_L t}
\right|^2,
\nonumber\\
\Gamma [ K^0 (t) \rightarrow \pi^+ e^- \nu ] &=&
\frac{| \langle \pi^+ e^- \nu | T | \overline{K^0} \rangle |^2}{4}
\left|
\left( \sqrt{1 - \theta^2} - \bar \lambda_e - \theta \bar \lambda_e \right)
e^{- i \mu_S t}
\right.
\nonumber\\
& &
\left.
-
\left( \sqrt{1 - \theta^2} + \bar \lambda_e - \theta \bar \lambda_e \right)
e^{- i \mu_L t} \right|^2
\frac{1 - \delta}{1 + \delta}.
\label{kaon decays}
\end{eqnarray}
These expressions should be compared with those used
in fitting the experimental data \cite{niebergall},
\begin{equation}
\Gamma [ K^0 (t) \rightarrow \pi^\mp e^\pm \nu ] \propto
\left| \left( 1 + x_e \right) e^{- i \mu_S t}
\pm \left( 1 - x_e \right) e^{- i \mu_L t} \right|^2.
\end{equation}
It is seen that the parameter $x_e$ to which the decay curves have been fitted
becomes ill-defined when one allows for CPT violation;
in one case one has $x_e \approx \theta - \lambda_e$,
in the other one it is $x_e \approx - \bar \lambda_e$.

Anyway,
we may state that the search for $\Delta S = - \Delta Q$ amplitudes
has provided a loose,
indirect bound on CPT violation in $K^0$-$\overline{K^0}$ mixing;
since one has obtained
$x_e$ \raise.3ex\hbox{$<$\kern-.75em\lower1ex\hbox{$\sim$}} $10^{-2}$,
one can also state that
$\theta$ \raise.3ex\hbox{$<$\kern-.75em\lower1ex\hbox{$\sim$}} $10^{-2}$.
Indeed,
if $\theta$ were much larger than this,
its effect should be visible in the analysis of the decay curves
in Eqs.~(\ref{kaon decays}).

Similarly,
the search for $\Delta B = - \Delta Q$ amplitudes
at the $\Upsilon(4S)$-factories has been discussed
assuming CPT invariance \cite{Dass94}.

\section{Decays from a correlated state}

Let us consider the correlated state with P- and C-parity $-1$ which,
at time $t=0$,
is
\begin{equation}
\phi^- = {\textstyle \frac{1}{\sqrt{2}}} \left[
| P^0 (\vec{k}) \rangle \, | \overline{P^0} (- \vec{k}) \rangle
-
| \overline{P^0} (\vec{k}) \rangle \, | P^0 (- \vec{k}) \rangle \right],
\label{phi-}
\end{equation}
where $\vec{k}$ and $- \vec{k}$ denote the opposite three-momenta
of the two mesons.
Consider what can be measured using the correlated state in Eq.~(\ref{phi-}).
Let the meson with momentum $\vec{k}$ decay at time $t_1$ into a state $f$,
and the meson with momentum $- \vec{k}$ decay at time $t_2$ into a state $g$.
Using Eqs.~(\ref{evolution}) one finds that
the decay amplitude may be written in the form
\begin{eqnarray}
\langle f, t_1; g, t_2 | T | \phi^- \rangle &=&
a \left[ g_+ (t_1) g_+ (t_2) - g_- (t_1) g_- (t_2) \right]
\nonumber\\ & &
+ b \left[  g_+ (t_1) g_- (t_2) - g_- (t_1) g_+ (t_2) \right],
\label{a and b}
\end{eqnarray}
where
\begin{eqnarray}
a &=& A_f \bar A_g - \bar A_f A_g,
\nonumber\\
b &=& \sqrt{1 - \theta^2}
\left( \frac{p}{q} A_f A_g - \frac{q}{p} \bar A_f \bar A_g \right)
+ \theta \left( A_f \bar A_g + \bar A_f A_g \right).
\end{eqnarray}
Since the decay amplitude in Eq.~(\ref{a and b})
depends on two independent time functions,
we may in principle extract the ratio of their coefficients:
\begin{equation}
F =  \frac{b}{a}
= \frac{\theta A_f \bar A_g + \theta \bar A_f A_g
+ \frac{p}{q} \sqrt{1 - \theta^2} A_f A_g
- \frac{q}{p} \sqrt{1 - \theta^2} \bar A_f \bar A_g}
{A_f \bar A_g - \bar A_f A_g}.
\label{measurable}
\end{equation}
Indeed,
by observing the shape of the dependence on $t_1$ and on $t_2$
of the decay rate,
\begin{eqnarray}
\left| \langle f, t_1; g, t_2 | T | \phi^- \rangle \right|^2
&=&
\left| a \right|^2 e^{- \Gamma \left( t_1 + t_2 \right)} \left\{
\frac{1 + \left| F \right|^2}{2}
\cosh \left[ \Gamma y \left( t_1 - t_2 \right) \right]
\right.
\nonumber\\
& & + \mbox{Re}\, F \sinh \left[ \Gamma y \left( t_1 - t_2 \right) \right]
- \mbox{Im}\, F \sin \left[ \Gamma x \left( t_1 - t_2 \right) \right]
\nonumber\\
& &
+ \left. \frac{1 - \left| F \right|^2}{2}
\cos \left[ \Gamma x \left( t_1 - t_2 \right) \right]
\right\},
\label{great decay rate}
\end{eqnarray}
one can determine $F$.
It turns out that,
as a matter of fact,
one can measure $F$ even if one integrates the decay rate over $t_1 + t_2$,
as long as one still follows its dependence on $t_1 - t_2$.
Indeed,
\begin{equation}
\int_{\left| t_1 - t_2 \right|}^{+ \infty} {\textstyle \frac{1}{2}}\,
d \left( t_1 + t_2 \right)
\left| \langle f, t_1; g, t_2 | T | \phi^- \rangle \right|^2
\end{equation}
is identical with the right-hand side
of Eq.~(\ref{great decay rate}) but for the overall factor,
which is $|a|^2 \exp \left( - \Gamma \left| t_1 - t_2 \right| \right)
/ \left( 2 \Gamma \right)$
instead of $|a|^2 \exp \left[ - \Gamma \left( t_1 + t_2 \right) \right]$.

\subsection{Di-lepton decays}

In the measurable quantity of Eq.~(\ref{measurable}),
let us suppose that $f$ is the semileptonic state $X^- l^+ \nu_l$,
while $g$ is another semileptonic state,
which has a charged lepton with the same electric charge,
$X^{\prime -} l^{\prime +} \nu_{l^\prime}$.
The quantity $F$ in Eq.~(\ref{measurable}) then reads
\begin{equation}
\frac{\theta x_{l^\prime} + \theta x_l
+ \frac{p}{q} \sqrt{1 - \theta^2}
- \frac{q}{p} \sqrt{1 - \theta^2} x_l x_{l^\prime}}
{x_{l^\prime} - x_l}
\approx \left( \lambda_{l^\prime} - \lambda_l \right)^{-1}.
\label{2lep-1}
\end{equation}
We have once again made the approximation of neglecting the products
of any two small parameters like $\theta$,
$\lambda_l$,
and $\lambda_{l^\prime}$.
Let us now suppose that $f$ is the semileptonic state $X^+ l^- \bar \nu_l$,
while $g$ is the semileptonic state
$X^{\prime +} l^{\prime -} \bar \nu_{l^\prime}$,
with a charged lepton with the same electric charge.
The quantity $F$ then reads
\begin{equation}
\frac{\theta \bar x_l^\ast + \theta \bar x_{l^\prime}^\ast
+ \frac{p}{q} \sqrt{1 - \theta^2} \bar x_l^\ast \bar x_{l^\prime}^\ast
- \frac{q}{p} \sqrt{1 - \theta^2}}
{\bar x_l^\ast - \bar x_{l^\prime}^\ast}
\approx \left( \bar \lambda_{l^\prime} - \bar \lambda_l \right)^{-1}.
\label{2lep-2}
\end{equation}
If,
instead,
the two semileptonic states detected in opposite sides of the detector
have charged leptons with opposite electric charge,
then the quantity in Eq.~(\ref{measurable}) is either
\begin{equation}
\frac{\theta + \theta x_l \bar x_{l^\prime}^\ast
+ \frac{p}{q} \sqrt{1 - \theta^2} \bar x_{l^\prime}^\ast
- \frac{q}{p} \sqrt{1 - \theta^2} x_l}
{1 - x_l \bar x_{l^\prime}^\ast}
\approx \theta + \bar \lambda_{l^\prime} - \lambda_l
\label{2lep-3}
\end{equation}
or,
with $l \leftrightarrow l^\prime$,
\begin{equation}
\frac{\theta + \theta x_{l^\prime} \bar x_l^\ast
+ \frac{p}{q} \sqrt{1 - \theta^2} \bar x_l^\ast
- \frac{q}{p} \sqrt{1 - \theta^2} x_{l^\prime}}
{1 - x_{l^\prime} \bar x_l^\ast}
\approx \theta + \bar \lambda_l - \lambda_{l^\prime}.
\label{2lep-4}
\end{equation}

We thus conclude that,
by using the decays of the correlated state $\phi^-$,
one can---at least in principle---measure
the four linear combinations of small parameters
\begin{eqnarray}
 & & \lambda_{l^\prime} - \lambda_l,
\nonumber\\
 & & \bar \lambda_{l^\prime} - \bar \lambda_l,
\nonumber\\
 & & \theta + \bar \lambda_{l^\prime} - \lambda_l,
\nonumber\\
 & & \theta + \bar \lambda_l - \lambda_{l^\prime}.
\end{eqnarray}
This means that {\em it is impossible to disentangle
the CPT-violating parameter $\theta$
from the $\Delta P = - \Delta Q$ parameters $\lambda_l$ and $\bar \lambda_l$
by using di-lepton events alone}.
This conclusion was also arrived at by Xing \cite{Xing98}.

\subsection{Single-lepton events}

Let us consider again the correlated state in Eq.~(\ref{phi-}).
We shall now study events in which the final state $f$
is observed in one side of the detector,
irrespectively of the decay occurring in the opposite side.
This corresponds to integrating over $t_2$
and summing over all final states $g$.
Using the unitarity conditions in Eqs.~(\ref{as primeiras})
and (\ref{a terceira}) of the Appendix,
one easily shows \cite{yamamoto} that
\begin{eqnarray}
\Gamma[\phi^- \rightarrow f](t_1)
& = &
\int_0^{+ \infty} \!dt_2 \sum_g
\left( \left| \langle f, t_1; g, t_2 | T | \phi^- \rangle \right|^2
+ \left| \langle g, t_2; f, t_1 | T | \phi^- \rangle \right|^2 \right)
\nonumber\\
& = &
\Gamma[P^0 (t_1) \rightarrow f] + \Gamma[\overline{P^0} (t_1) \rightarrow f],
\label{yama}
\end{eqnarray}
where $\Gamma[P^0 (t_1) \rightarrow f]$ is the rate for a single,
tagged $P^0$ to decay into the final state $f$ at time $t_1$,
while $\Gamma[\overline{P^0} (t_1) \rightarrow f]$ is the decay rate
for a single tagged $\overline{P^0}$.
These rates have been given in Eqs.~(\ref{tagged decay rates}).
One finds
\begin{eqnarray}
\Gamma [\phi^- \rightarrow f] (t) &=&
\left| g_+(t) \right|^2 \left( |A_f|^2 + |\bar A_f|^2 \right)
\nonumber\\ 
& & + \left| g_-(t) \right|^2 \left\{ \left( \left| \theta \right|^2
+ \left| 1 - \theta^2 \right| \frac{1 + \delta^2}{1 - \delta^2} \right)
\left( |A_f|^2 + |\bar A_f|^2 \right) \right.
\nonumber\\ 
& & + \left| 1 - \theta^2 \right| \frac{2 \delta}{1 - \delta^2}
\left( |A_f|^2 - |\bar A_f|^2 \right)
\nonumber\\ 
& & \left. + 2\, \mbox{Re} \left[ \theta^\ast \sqrt{1 - \theta^2}
\left( \frac{p}{q} A_f \bar A_f^\ast
- \frac{q}{p} A_f^\ast \bar A_f \right) \right] \right\}
\nonumber\\ 
& & + 2\, \mbox{Re} \left\{ g_+^\ast (t) g_-(t)
\left[ \theta \left( |\bar A_f|^2 - |A_f|^2 \right) \right. \right.
\nonumber\\ & & \left. \left. + \sqrt{1 - \theta^2}
\left( \frac{q}{p} A_f^\ast \bar A_f + \frac{p}{q} A_f \bar A_f^\ast \right)
\right] \right\}.
\label{the general case}
\end{eqnarray}

Consider the particular case
in which $f$ is the semileptonic state $X^- l^+ \nu_l$.
The small parameters are $\delta$
(which violates T and CP),
$\theta$
(which violates CPT and CP),
and $\lambda_l$
(which violates the $\Delta P = \Delta Q$ rule).
Working out Eq.~(\ref{the general case})
to subleading order in those parameters,
one finds
\begin{eqnarray}
\frac{\Gamma \left[ \phi^- \rightarrow X^- l^+ \nu_l \right] (t)}
{| \langle X^- l^+ \nu_l | T | P^0 \rangle |^2}
 & \approx &
\left| g_+(t) \right|^2 + \left| g_-(t) \right|^2 \left( 1 + 2 \delta \right)
\nonumber\\ 
& & + 2\, \mbox{Re} \left[ g_+^\ast(t) g_-(t) \right]
\left[ - \mbox{Re}\, \theta + 2 
\left( 1 + \delta \right) \mbox{Re}\, \lambda_l \right]
\nonumber\\ 
& & + 2\, \mbox{Im} \left[ g_+^\ast(t) g_-(t) \right]
\left( \mbox{Im}\, \theta + 2 \delta \mbox{Im}\, \lambda_l \right).
\label{na segunda aprox}
\end{eqnarray}
Let us instead take $f$
to be the semileptonic final state $X^+ l^- \bar \nu_l$.
We find
\begin{eqnarray}
\frac{\Gamma \left[ \phi^- \rightarrow X^+ l^- \bar \nu_l \right] (t)}
{| \langle X^+ l^- \bar \nu_l | T | \overline{P^0} \rangle |^2}
& \approx & \left| g_+(t) \right|^2 + \left| g_-(t) \right|^2
\left( 1 - 2 \delta \right)
\nonumber\\ & & + 2\, \mbox{Re} \left[ g_+^\ast(t) g_-(t) \right]
\left[ \mbox{Re}\, \theta
+ 2 \left( 1 - \delta \right) \mbox{Re}\, \bar \lambda_l \right]
\nonumber\\ & & - 2\, \mbox{Im} \left[ g_+^\ast(t) g_-(t) \right]
\left( \mbox{Im}\, \theta + 2  \delta \mbox{Im}\, \bar \lambda_l \right).
\end{eqnarray}
The functions of time $\left| g_+(t) \right|^2$,
$\left| g_-(t) \right|^2$,
$\mbox{Re} \left[ g_+^\ast(t) g_-(t) \right]$,
and $\mbox{Im} \left[ g_+^\ast(t) g_-(t) \right]$ are independent.
Therefore,
the three ratios of their coefficients may in principle
be extracted from experiment.
One sees that,
if one neglects the subleading terms,
then the coefficients of the time function
$\mbox{Im} \left[ g_+^\ast(t) g_-(t) \right]$
yield the CPT-violating parameter $\mbox{Im}\, \theta$.
On the other hand,
$\mbox{Re}\, \theta$ cannot,
even in this approximation,
be separated from $\mbox{Re}\, \lambda_l$
or $\mbox{Re}\, \bar \lambda_l$.

If one does not neglect the subleading terms,
then one can determine the T-violating parameter $\delta$
from the coefficients of the function $\left| g_- (t) \right|^2$.
On the other hand,
even if one measures a non-zero coefficient of
$\mbox{Im} \left[ g_+^\ast(t) g_-(t) \right]$,
one cannot ascertain that one has found CPT violation.
This is because of the subleading terms
$\delta \mbox{Im}\, \lambda_l$ or $\delta \mbox{Im}\, \bar \lambda_l$,
which are CPT-invariant.

\subsection{Combining single-lepton and di-lepton events}

So far,
we have separately discussed the impact that 
CPT violation and wrong-charge semileptonic decays have on correlated
decays into two semileptonic and into one semileptonic final state.
Schematically,
we have found that
\begin{eqnarray}
\label{phi-:l+l-}
\phi^- \rightarrow l^+ l^-
&\Longrightarrow& 
\left\{ \begin{array}{l}
   \mbox{Re}\, \theta + \mbox{Re}\, \bar \lambda_l - \mbox{Re}\, \lambda_l\\
   \mbox{Im}\, \theta + \mbox{Im}\, \bar \lambda_l - \mbox{Im}\, \lambda_l
	\end{array}
\right.,
\\
\label{phi-:l+}
\phi^- \rightarrow l^+
&\Longrightarrow& 
\left\{ \begin{array}{l}
	2 \mbox{Re}\, \lambda_l - \mbox{Re}\, \theta\\
	2 \delta \mbox{Im}\, \lambda_l + \mbox{Im}\, \theta
	\end{array}
\right.,
\\
\label{phi-:l-}
\phi^- \rightarrow l^-
&\Longrightarrow& 
\left\{ \begin{array}{l}
	2 \mbox{Re}\, \bar \lambda_l + \mbox{Re}\, \theta\\
	2 \delta \mbox{Im}\, \bar \lambda_l + \mbox{Im}\, \theta
	\end{array}
\right..
\end{eqnarray}
Combining the first lines of Eqs.~(\ref{phi-:l+}) and (\ref{phi-:l-})
we may extract
$2 \mbox{Re} \left( \theta + \bar \lambda_l -  \lambda_l \right)$,
which is the same information as in the first line
of Eq.~(\ref{phi-:l+l-}).
Thus,
{\em $\mbox{Re}\, \theta$ cannot be disentangled from
$\mbox{Re} \left( \bar \lambda_l - \lambda_l \right)$}.

On the other hand,
one can combine the di-lepton and single-lepton events to extract
$\mbox{Im}\, \theta$.
In fact,
we may combine the second lines of Eqs.~(\ref{phi-:l+}) and (\ref{phi-:l-})
to extract $\delta \mbox{Im} \left( \bar \lambda_l - \lambda_l \right)$.
This yields $\mbox{Im} \left( \bar \lambda_l - \lambda_l \right)$,
provided one is able to determine $\delta$
from the coefficients of $\left| g_- (t) \right|^2$
in the single-lepton events.
Then,
by comparing $\mbox{Im} \left( \bar \lambda_l - \lambda_l \right)$
with the second line of Eq.~(\ref{phi-:l+l-}),
one can determine $\mbox{Im}\, \theta$ unambiguously.

We conclude that the decays of the correlated state $| \phi^- \rangle$ can,
in principle,
be used to disentangle violations of CPT from violations of the
$\Delta P = \Delta Q$ rule in semileptonic decays.
However,
this only happens in the imaginary parts of the parameters;
the real parts remain un-separated,
even if one takes into account both di-lepton
and single-lepton events.

It is important to observe that this determination of $\mbox{Im}\, \theta$
is possible even in the case $\Gamma_a = \Gamma_b$,
which is expected to hold to a good approximation
in the $B^0_d$-$\overline{B^0_d}$ system.
Indeed,
in that case the measurement of
the first lines of Eqs.~(\ref{phi-:l+l-})--(\ref{phi-:l-})
will be impossible,
but the measurement of their second lines will still be feasible.
Thus,
the extraction of $\mbox{Im}\, \theta$ will not be impeded
by an approximate equality
of the decay widths of the two eigenstates of mixing.

\section{New-physics effects in the production mechanism}

We have shown in section~3 that one may in principle
separate the CPT-violating parameter $\theta$
from the $\Delta P = - \Delta Q$ parameters $\lambda_l$
and $\bar \lambda_l$
by following the time dependence of the semileptonic decays
of single tagged mesons.
We recall that by ``tagged meson'' we mean a neutral meson
whose flavor has been unequivocally determined at time $t=0$.
This is normally done by evoking the rule of associated production,
which is based on the flavor-conserving nature
of the interactions of the gluon,
photon,
or $Z^0$, that are responsible for most production mechanisms.
It states that,
if an initial state $i$ decays into a neutral meson
together with a tagging state $n$,
then the flavour of $n$ is opposite to the one of the neutral meson.

To be specific,
let us consider the conditions at CPLEAR.
There one starts with $i = p \bar p$
and one looks for the charge of the kaon in the state $n =  K^- \pi^+$.
Since the strong interaction preserves flavor,
this will identify as $K^0$ the neutral meson produced in association with $n$.
Conversely,
if the charged kaon has positive charge,
then $n =  K^+ \pi^-$ and the neutral meson is $\overline{K^0}$.
In reality,
there could be a small $\left| \Delta S \right| = 2$ interaction
enabling the production process $p \bar p \rightarrow  K^+ \pi^- K^0$.
This would destroy the rule of associated production,
and one would lose the notion of tagged decays.
However,
that $\left| \Delta S \right| = 2$ effect would also contribute
to $K^0$-$\overline{K^0}$ mixing and,
thus,
one would expect it to be negligibly small.

On the contrary,
new-physics effects in the production mechanism may be important
when the production is due to the interaction
of the $W$ boson.
Here we should consider two cases,
depending on whether the leading-order SM production mechanism
allows for only one (say, $P^0$) or both ($P^0$ and $\overline{P^0}$)
neutral mesons to be produced in the final state.

The decay $B^0_d \rightarrow J/\psi K^0$
constitutes an example of the first case.
To leading order in the SM this decay takes place,
while $B^0_d \rightarrow J/\psi \overline{K^0}$ does not.
(These processes are called ``semileptonic-type decays''
by Kosteleck\'y and collaborators \cite{kostelecky}.)
In principle one could use the subsequent evolution of the neutral kaon
in order to test its properties;
in practice,
since those properties are rather well tested
in direct decays of neutral kaons,
these cascade decays should be used instead to probe the properties
of the $B^0_d$-$\overline{B^0_d}$ system \cite{cascade}. 
In this particular case,
the initial state can evolve into its CP conjugate.
We shall not consider such cases any further,
but we stress that even a small new-physics amplitude to the process 
$B^0_d \rightarrow J/\psi \overline{K^0}$ will affect those analyses.

There are also cascade decays
in which both neutral mesons can arise in the intermediate state,
even within the SM.
For example,
one may want to use the copious production of $B^-$ at the $\Upsilon(4S)$,
together with a subsequent decay $B^- \rightarrow K^- D^0$,
in order to probe the decays of the $D^0$ into some final state $f$. 
However,
in this case there is another possibility.
Although suppressed by about an order of magnitude in amplitude,
the process $B^- \rightarrow K^- \overline{D^0}$ is also allowed in the SM.
Thus,
if both $D^0 \rightarrow f$ and $\overline{D^0} \rightarrow f$ are allowed,
there are two interfering decay paths:
$B^- \rightarrow K^- D^0 \rightarrow K^- f$
and $B^- \rightarrow K^- \overline{D^0} \rightarrow K^- f$.

This is actually at the root of the Gronau--London--Wyler \cite{GLW}
and Atwood--Dunietz--Soni \cite{ADS} methods
to determine the CP-violating phase $\gamma$.
In those methods
one assumes that there is no $D^0$-$\overline{D^0}$ mixing;
in that case,
there is no interference effect for those final states $f$
into which either $\overline{D^0}$ or $D^0$ cannot decay.
Recently,
Meca and Silva \cite{Meca98}
have shown that the presence of $D^0$-$\overline{D^0}$ mixing
gives rise to a new interference effect,
between the amplitudes of the decays into the $D^0$-$\overline{D^0}$ system,
on the one hand,
and the mixing in that system,
on the other hand.
One of the consequences of this result is that,
even if we look for a final state $f$
into which $\overline{D^0}$ cannot decay,
there will be two interfering paths:
the unmixed decay path $B^- \rightarrow K^- D^0 \rightarrow K^- f$;
and the mixed decay path
$B^- \rightarrow K^- \overline{D^0} \rightarrow K^- D^0 \rightarrow K^- f$.
This effect makes it possible to test new sources of CP violation,
and might provide a handle on $\Delta m$
in the $D^0$-$\overline{D^0}$ system \cite{Meca98,Amorim99}.

We shall consider the general situation in which the initial state
$i$ can lead both to the state $n$ together with $P^0$,
and to the state $n$ together with $\overline{P^0}$.
That is,
we assume that
\begin{equation}
c_i = \langle n P^0 | T | i \rangle
\ \ \mbox{and} \ \
\bar c_i = \langle n \overline{P^0} | T | i \rangle
\end{equation}
are both non-vanishing.
Then,
the production process
leads to the superposition of $P^0$ and $\overline{P^0}$ given by
\begin{equation}
| \psi_i \rangle
= c_i | P^0 \rangle + \bar c_i | \overline{P^0} \rangle. 
\end{equation}
Using Eqs.~(\ref{evolution}),
we find that this state evolves into
\begin{eqnarray}
| \psi_i (t) \rangle &=& g_+ (t)
\left( c_i | P^0 \rangle + \bar c_i | \overline{P^0} \rangle \right)
+ g_- (t) \left[ c_i \left(
\frac{q}{p} \sqrt{1 - \theta^2} | \overline{P^0} \rangle
- \theta | P^0 \rangle \right) \right.
\nonumber\\
& & \left. + \bar c_i \left( \frac{p}{q} \sqrt{1 - \theta^2} | P^0 \rangle
+ \theta | \overline{P^0} \rangle \right) \right]
\end{eqnarray}
at time $t$.
Suppose that one observes experimentally
the time dependence of the overall process
\begin{equation}
i \rightarrow n \left\{ P^0, \overline{P^0} \right\} \rightarrow n f.
\end{equation}
Recalling that $g_+(t)$ and $g_-(t)$ are independent functions,
we conclude that
this allows in principle the determination of
\begin{equation}
\hat E =
\frac{ c_i \left( \frac{q}{p} \sqrt{1 - \theta^2} \bar A_f - \theta A_f \right)
+ \bar c_i
\left( \frac{p}{q} \sqrt{1 - \theta^2} A_f + \theta \bar A_f \right)}{
c_i A_f + \bar c_i \bar A_f}.
\end{equation}
Clearly,
if the state $n$ correctly tags $P^0$,
{\it i.e.},
if $\bar c_i = 0$,
then $\hat E$ coincides with $E$ in Eq.~(\ref{tagged measurables}).
Conversely,
if $n$ really identifies $\overline{P^0}$,
{\it i.e.},
if $c_i = 0$,
then $\hat E = \bar E$.

Let us define \cite{Meca98,Amorim99}
\begin{equation}
\xi_i = \frac{\bar c_i}{c_i}\, \frac{p}{q}
= \frac{\langle n \overline{P^0} | T | i \rangle}
{\langle n P^0 | T | i \rangle}\, \frac{p}{q}
\end{equation}
and $\bar \xi_i = 1 / \xi_i$.
The parameter $\xi_i$ describes the interference
between the production process,
represented by the two amplitudes $c_i$ and $\bar c_i$,
and the subsequent $P^0$-$\overline{P^0}$ mixing,
described by $q/p$.
We want to consider cases in which,
although $n$ is not a perfect tag for $P^0$
(or for $\overline{P^0}$),
the mis-tagging is small.
Then,
we may treat $\xi_i$ (or $\bar \xi_i$) as a small parameter.

When $f$ is the semileptonic state $X^- l^+ \nu_l$,
we may in principle measure,
if $\xi_i$ is small,
\begin{equation}
\hat E = \frac{\sqrt{1 - \theta^2} \lambda_l - \theta
+ \sqrt{1 - \theta^2} \xi_i + \theta \xi_i \lambda_l}
{1 + \xi_i \lambda_l}
\approx \lambda_l - \theta + \xi_i;
\label{vap-l+}
\end{equation}
or,
if $\bar \xi_i$ is small,
\begin{equation}
\hat E =
\frac{\sqrt{1 - \theta^2} \bar \xi_i \lambda_l - \theta \bar \xi_i
+ \sqrt{1 - \theta^2} + \theta \lambda_l}{\bar \xi_i + \lambda_l}
\approx \left( \bar \xi_i + \lambda_l \right)^{-1}.
\end{equation}
If $f = X^+ l^- \bar \nu_l$,
then we may in principle measure,
if $\xi_i$ is small,
\begin{equation}
\hat E =
\frac{\sqrt{1 - \theta^2} - \theta \bar \lambda_l
+ \sqrt{1 - \theta^2} \xi_i \bar \lambda_l + \theta \xi_i}
{\bar \lambda_l + \xi_i}
\approx \left( \bar \lambda_l + \xi_i \right)^{-1};
\end{equation}
or,
if $\bar \xi_i$ is small,
\begin{equation}
\hat E =
\frac{\sqrt{1 - \theta^2} \bar \xi_i - \theta \bar \xi_i \bar \lambda_l
+ \sqrt{1 - \theta^2} \bar \lambda_l + \theta}
{\bar \xi_i \bar \lambda_l + 1}
\approx \bar \xi_i + \bar \lambda_l + \theta
\label{vap-l-}.
\end{equation}
In all cases we have made the approximation of neglecting products of
small parameters.

Equations~(\ref{vap-l+})--(\ref{vap-l-}) should be compared with
Eqs.~(\ref{tagged-l+}) and (\ref{tagged-l-}).
We see that,
if both neutral mesons can be produced in association with $n$,
then we may no longer disentangle the CPT-violating parameter
$\theta$ from the effects of both mis-tagging
and $\Delta P = - \Delta Q$ amplitudes,

Notice that we have only considered cases in which the initial
state $i$ cannot evolve into $\bar i$.
This includes cascade decays originating in baryons or charged mesons,
but not cascade decays which start from
a heavier neutral-meson-antimeson system.
In the latter cases
the analysis is more complicated \cite{cascade,Amorim99}
because there are two distinct neutral-meson systems evolving in time.

\subsection{Di-lepton decays and mis-tagging}

It is instructive to view the difficulties with di-lepton decays,
discussed in section~4,
as a particular case of mis-tagging.
Let us review the tagging strategy usually evoked for
measurements at the $\Upsilon(4S)$.
It is generally assumed that there are no violations of the
$\Delta B = \Delta Q$ rule in semileptonic decays.
Then the following reasoning applies:
1) although the $B^0_d$ and the $\overline{B^0_d}$
produced at the $\Upsilon(4S)$ oscillate,
the antisymmetry of the wave function
is preserved by the linearity of the evolution;
2) hence,
if at some instant the right-moving meson is found
(from its semileptonic decay) to be $B^0_d$,
then the left-moving meson at that instant is certainly $\overline{B^0_d}$;
3) that left-moving meson will evolve
from that instant onwards as a tagged $\overline{B^0_d}$;
4) therefore,
time-dependent experiments starting from the state $\Upsilon(4S)$
and observing at least one semileptonic decay automatically reproduce
the results obtained with tagged decays.

If we allow for violations of the $\Delta B = \Delta Q$ rule,
then the situation obtained,
in Eqs.~(\ref{2lep-1})--(\ref{2lep-4}),
is the same as the one
in the right-hand-sides of Eqs.~(\ref{vap-l+})--(\ref{vap-l-}),
but with the substitutions
$\xi_i = - \bar \lambda_{l^\prime}$ and $\bar \xi_i = - \lambda_{l^\prime}$.
The reason is simple:
since we are assuming violations of the $\Delta B = \Delta Q$ rule,
the semileptonic decays do not provide a perfect tagging
of the neutral mesons originated from the $\Upsilon(4S)$.
If the right-moving meson decays at some instant
into $X^{\prime +} l^{\prime -} \bar \nu_{l^\prime}$,
then we know that the left-moving meson is,
at that instant,
in a state which has zero probability of decaying
into $X^{\prime +} l^{\prime -} \bar \nu_{l^\prime}$;
that state is
\begin{equation}
\langle X^{\prime +} l^{\prime -} \bar \nu_{l^\prime}
| T | \overline{P^0} \rangle | P^0 \rangle
- \langle X^{\prime +} l^{\prime -} \bar \nu_{l^\prime}
| T | P^0 \rangle | \overline{P^0} \rangle.
\end{equation}
If one observes the decay of that left-moving meson
into any other state,
at any other time,
then we are effectively working with a (mis)tagged state having
\begin{equation}
\xi_i = \frac{p}{q}
\left[ - \frac{\langle X^{\prime +} l^{\prime -} \bar \nu_{l^\prime}
| T | P^0 \rangle}
{\langle X^{\prime +} l^{\prime -} \bar \nu_{l^\prime}
| T | \overline{P^0} \rangle} \right]
= - \bar \lambda_{l^\prime}.
\end{equation}
The decays of the $\Upsilon(4S)$ into one semileptonic state and another
final state $f$ thus become equivalent to mis-tagged decays,
{\it i.e.},
to decays in which the tagging strategy does not work properly.

This analysis has a very important implication:
{\em the case of tagged decays}
(which would allow us
to extract unambiguously the CPT-violating parameter $\theta$)
{\em can never be implemented at the $\Upsilon(4S)$,
if one allows for the existence of $\Delta P \neq \Delta Q$ amplitudes}.

The same ``no-go theorem''
applies to almost all existing or proposed $B$-physics experiments.
Indeed,
one looks almost always for the decay of 
the mesons produced in association
with the neutral $B$ meson which one wishes to tag,
and that decay may also be affected by new physics.
This is clearly the case for correlated or uncorrelated
$B^0 \overline{B^0}$ production,
but it also occurs for the production of $B^- B^0 X^+$,
if one only identifies the $B^-$
through the sign of the lepton in the final state.
The exception occurs in the case of $B^- B^0 X^+$ production,
if one detects the $B^-$ by a full reconstruction of the event.
Although inefficient,
this method guarantees, in principle,
that the tagging meson is indeed a $B^-$ and thus,
barring new effects in the production mechanism,
it identifies the neutral meson at the time of production as $B^0$.

We should point out that we have not discussed the possibility
that there might be new $\left| \Delta B \right| = 2$ effects
in the decay of the $\Upsilon(4S)$.
In this case,
the $B^0_d \overline{B^0_d}$ wave function
would no longer have the form in Eq.~(\ref{phi-});
it would also have $B^0_d B^0_d$
and $\overline{B^0_d}\, \overline{B^0_d}$ components.
This possibility was hinted at
(but not explored)
by Yamamoto in the first article of \cite{yamamoto};
he pointed out that it would invalidate Eq.~(\ref{yama}).

\section{Conclusions}

In this article we have discussed ways
to uncover a signal of CPT violation using semileptonic decays.
We have stressed the fact that such a signal must be disentangled
from new-physics effects in the tagging procedure.
These can be due to $\Delta P = - \Delta Q$ amplitudes
in semileptonic decays;
but also to flavor-conservation violations in the production mechanism.

If one assumes the latter to be absent,
then one concludes that:
\begin{enumerate}
\item The separation between violations of CPT
and violations of the $\Delta P = \Delta Q$ rule
is possible using tagged decays;
unfortunately,
these studies cannot be implemented at the $\Upsilon(4S)$
and are difficult to implement elsewhere.
\item One cannot disentangle violations of the $\Delta P = \Delta Q$ rule
from violations of CPT using di-lepton decays of the state $| \phi^- \rangle$.
\item This can be achieved if one uses both di-lepton
and single-lepton decays,
but even then one can only determine the CPT-violating parameter
$\mbox{Im}\, \theta$,
while $\mbox{Re}\, \theta$ remains unknown.
\end{enumerate}

However,
one cannot exclude the possibility that there are also
new interactions in the production process.
Such effects may destroy the notion of ``tagged decays''.
They are expected to be negligible when the production mechanism
is mediated by gluons,
photons,
or by the $Z^0$ boson;
the rule of associated production
should then hold to sufficient accuracy.
However,
if that is not the case
(as, for example, in cascade decays),
then the clear identification of CPT violation 
using only semileptonic decays becomes impossible.

\section*{Appendix}

In this appendix we discuss the unitarity conditions.
They are needed in order to prove Eq.~(\ref{yama}).
We start from the equation
\begin{equation}
- \frac{d}{dt} \langle \psi (t) | \psi (t) \rangle
= \sum_g \left| \langle g | T | \psi (t) \rangle \right|^2,
\label{probability conservation}
\end{equation}
which expresses the conservation of probability
in the decay of the state in Eq.~(\ref{the main state}).
Using Eq.~(\ref{the main evolution}) for arbitrary values
of $\psi_1 (t)$ and $\psi_2 (t)$,
one finds that Eq.~(\ref{probability conservation}) yields
\begin{eqnarray}
\sum_g \left| A_g \right|^2 &=& - 2\, \mbox{Im}\, R_{11},
\nonumber\\
\sum_g \left| \bar A_g \right|^2 &=& - 2\, \mbox{Im}\, R_{22},
\nonumber\\
\sum_g A_g^\ast \bar A_g &=& i \left( R_{12} - R_{21}^\ast \right).
\end{eqnarray}
We now express the matrix elements of $R$
in terms of the physical parameters $\mu_a$,
$\mu_b$,
$\theta$,
and $\delta$.
We thus obtain the unitarity conditions in the presence of violations of CPT:
\begin{eqnarray}
\sum_g \left| A_g \right|^2 &=& \Gamma
\left( 1 + x\, \mbox{Im}\, \theta - y\, \mbox{Re}\, \theta \right),
\nonumber\\
\sum_g \left| \bar A_g \right|^2 &=& \Gamma
\left( 1 - x\, \mbox{Im}\, \theta + y\, \mbox{Re}\, \theta \right),
\nonumber\\
\sum_g \frac{q}{p} A_g^\ast \bar A_g &=& \Gamma
\frac{\left( y + i \delta x \right) \mbox{Re} \sqrt{1 - \theta^2}
- \left( x - i \delta y \right) \mbox{Im} \sqrt{1 - \theta^2}}{1 + \delta}.
\label{unitarity}
\end{eqnarray}
In Eqs.~(\ref{probability conservation})--(\ref{unitarity})
the sums run over all the available decay modes $g$.

Kenny and Sachs \cite{kenny} have questioned the use of some simpler
unitarity conditions when testing CPT invariance,
on the grounds that one of the assumptions of the CPT theorem
is the hermiticity of the interactions.
However,
our derivation of the unitarity conditions is directly rooted
on the conservation of probability
expressed by Eq.~(\ref{probability conservation}),
and not on the hermiticity of the Hamiltonian.
We would agree with Tanner and Dalitz \cite{Tanner86},
who argue that any theory which violates Eq.~(\ref{probability conservation})
should probably be regarded as physically unacceptable.

We need the following integrals:
\begin{eqnarray}
\int_0^{+ \infty} \!dt \left| g_+ (t) \right|^2 &=& \frac{2 + x^2 - y^2}
{2 \Gamma \left( 1 - y^2 \right) \left( 1 + x^2 \right)},
\nonumber\\
\int_0^{+ \infty} \!dt \left| g_- (t) \right|^2 &=& \frac{x^2 + y^2}
{2 \Gamma \left( 1 - y^2 \right) \left( 1 + x^2 \right)},
\nonumber\\
\int_0^{+ \infty} \!dt\, g_+^\ast (t) g_- (t) &=&
\frac{- y \left( 1 + x^2 \right) - i x \left( 1 - y^2 \right)}
{2 \Gamma \left( 1 - y^2 \right) \left( 1 + x^2 \right)}.
\label{integrated g-squares}
\end{eqnarray}

Remembering Eqs.~(\ref{evolution}),
one may use Eqs.~(\ref{unitarity}) and (\ref{integrated g-squares})
to show that
\begin{eqnarray}
\int_0^{+ \infty} \!dt\, \sum_g
\left| \langle g | T | P^0 (t) \rangle \right|^2
& = &
1,
\nonumber\\
\int_0^{+ \infty} \!dt\, \sum_g
\left| \langle g | T | \overline{P^0} (t) \rangle \right|^2
& = &
1,
\label{as primeiras}
\end{eqnarray}
as one would expect.
Also,
\begin{equation}
\int_0^{+ \infty} \!dt\, \sum_g \langle g | T | P^0 (t) \rangle \,
\langle g | T | \overline{P^0} (t) \rangle^\ast
= 0.
\label{a terceira}
\end{equation}

Equation~(\ref{yama}) follows from Eqs.~(\ref{as primeiras})
and (\ref{a terceira}).
A result similar to Eq.~(\ref{yama}),
but with $\phi^-$ substituted by the state with P- and C-parity $+1$,
also holds \cite{yamamoto},
as one would expect on the basic grounds of the conservation of probability.


\begin{thebibliography}{99}
%
\bibitem{proofs}
R.\ Jost,
Helv.\ Phys.\ Acta {\bf 30}, 409 (1957).
For recent presentations,
see for example R.\ F.\ Streater and A.\ S.\ Wightman,
{\it ``PCT, Spin, Statistics and All That''},
Benjamin, New York (1968);
S.\ Weinberg,
{\it ``The Quantum Theory of Fields''},
Cambridge University Press (1995).
%
\bibitem{PDG}
Particle Data Group (C.\ Caso {\it et al.}),
Eur.\ Phys.\ J.\ C {\bf 3}, 1 (1998).
%
\bibitem{Chr64}
J.\ H.\ Christenson, J.\ W.\ Cronin,
V.\ L.\ Fitch, and R.\ Turlay,
Phys.\ Rev.\ Lett.\ {\bf 13}, 138 (1964).
%
\bibitem{CPLEAR-T}
CPLEAR Collaboration (A.\ Angelopoulos {\it et al.}),
Phys.\ Lett.\ B {\bf 444}, 43 (1998).
%
\bibitem{CPLEAR-CPT}
CPLEAR Collaboration (A.\ Angelopoulos {\it et al.}),
Phys.\ Lett.\ B {\bf 444}, 52 (1998).
%
\bibitem{OPAL}
OPAL Collaboration (K.\ Ackerstaff {\it et al.}),
Z.\ Phys.\ C {\bf 76}, 401 (1997). 
%
\bibitem{cascade}
Ya.\ I.\ Azimov, JETP Letters {\bf 50}, 447 (1989);
Phys.\ Rev.\ D {\bf 42}, 3705 (1990);
G.\ V.\ Dass and K.\ V.\ L.\ Sarma,
Int.\ J.\  Mod.\ Phys.\ A {\bf 7}, 6081 (1992);
erratum {\it ibid} {\bf A8}, 1183 (1993);
B.\ Kayser and L.\ Stodolsky,
Max-Planck Institute report MPI-PHT-96-112 (1996),
unpublished
(hep-ph/9610522);
B.\ Kayser, in {\it ``Les Arcs 1997, Electroweak
interactions and unified theories''}, edited by J.\ Tran Thanh Van,
Editions Fronti\`eres, Paris (1997);
Ya.\ I.\ Azimov and I.\ Dunietz,
Phys.\ Lett.\ B {\bf 395}, 334 (1997).
%
\bibitem{Meca98}
C.\ C.\ Meca and J.\ P.\ Silva,
Phys.\ Rev.\ Lett.\ {\bf 81}, 1377 (1998).
%
\bibitem{Amorim99}
A.\ Amorim, M.\ G.\ Santos, and J.\ P.\ Silva,
Phys.\ Rev.\ D {\bf 59}, 056001 (1999).
%
\bibitem{Sachs63}
R.\ G.\ Sachs,
Phys.\ Rev.\ {\bf 129}, 2280 (1963);
Prog.\ Theor.\ Phys.\ {\bf 54}, 809 (1975).
%
\bibitem{Dass72}
G.\ V.\ Dass and P.\ K.\ Kabir,
Proc.\ Roy.\ Soc.\ London A {\bf 330}, 331 (1972).
%
\bibitem{Barmin84}
V.\ V.\ Barmin {\it et al.},
Nucl.\ Phys.\ B {\bf 247}, 293 (1984).
%
\bibitem{Tanner86}
N.\ W.\ Tanner and R.\ H.\ Dalitz,
Ann.\ Phys.\ (N.Y.) {\bf 171}, 463 (1986).
%
\bibitem{Lavoura91}
L.\ Lavoura,
Ann.\ Phys.\ (N.Y.) {\bf 207}, 428 (1991).
%
\bibitem{Gaume98}
L.\ Alvarez-Gaum\'{e}, C.\ Kounnas, S.\ Lola, and P.\ Pavlopoulos,
CERN report CERN-TH/98-392 (1998), 
unpublished (hep-ph/9812326).
%
\bibitem{Kabir99}
P.\ K.\ Kabir,
University of Virginia report UVA/99-137 (1999),
unpublished (hep-ph/9901340).
%
\bibitem{Briere89}
R.\ A.\ Briere and L.\ H.\ Orr,
Phys.\ Rev.\ D {\bf 40}, 2269 (1989).
%
\bibitem{Kobayashi92}
M.\ Kobayashi and A.\ I.\ Sanda,
Phys.\ Rev.\ Lett.\ {\bf 69}, 3139 (1992).
%
\bibitem{Xing94}
Z.-Z.\ Xing,
Phys.\ Rev.\ D {\bf 50}, R2957 (1994).
%
\bibitem{Mohapatra98}
A.\ Mohapatra, M.\ Satpathy, K.\ Abe, and Y.\ Sakay,
Phys.\ Rev.\ D {\bf 58}, 036003 (1998).
%
\bibitem{kostelecky}
V.\ A.\ Kosteleck\'{y} and R.\ Potting,
Phys.\ Rev.\ D {\bf 51}, 3923 (1995);
D.\ Colladay and V.\ A.\ Kosteleck\'{y},
Phys.\ Lett.\ B {\bf 344}, 259 (1995);
V.\ A.\ Kosteleck\'{y} and R.\ Van Kooten,
Phys.\ Rev.\ D {\bf 54}, 5585 (1996).
%
\bibitem{niebergall}
F.\ Niebergall {\it et al.},
Phys.\ Lett.\ {\bf 49B}, 103 (1974).
For a complete list of references,
see \cite{PDG}, pp.\ 469--471.
%
\bibitem{Dass94}
G.\ V.\ Dass and K.\ V.\ L.\ Sarma,
Phys.\ Rev.\ Lett.\ {\bf 72}, 191 (1994);
erratum {\it ibid.} {\bf 72}, 1573 (1994);
Phys.\ Rev.\ D {\bf 54}, 5880 (1996);
Tata Institute of Fundamental Research
report TIFR/TH/96-57 (1996),
unpublished (hep-ph/9610466).
%
\bibitem{Xing98}
Z.-Z.\ Xing,
University of Munich report LMU-98-11 (1998),
unpublished (hep-ph/9810249).
%
\bibitem{yamamoto}
H.\ Yamamoto,
Phys.\ Rev.\ Lett.\ {\bf 79}, 2402 (1997);
Phys.\ Lett.\ B {\bf 401}, 91 (1997).
%
\bibitem{GLW}
M.\ Gronau and D.\ London,
Phys.\ Lett.\ B {\bf 253}, 483 (1991);
M.\ Gronau and D.\ Wyler,
Phys.\ Lett.\ B {\bf 265}, 172 (1991).
%
\bibitem{ADS}
D.\ Atwood, I.\ Dunietz, and A.\ Soni,
Phys.\ Rev.\ Lett.\ {\bf 78}, 3257 (1997).
%
\bibitem{kenny}
B.\ G.\ Kenny and R.\ G.\ Sachs,
Phys.\ Rev.\ D {\bf 8}, 1605 (1973).
%
\end{thebibliography}
\end{document}